\documentclass[showpacs,preprintnumbers,amsmath,amssymb,      
twocolumn, tightenlines,prl
]{revtex4}

\usepackage[dvips]{graphicx}
\usepackage{bm}
\usepackage{amsmath}
\usepackage{dcolumn}

\begin{document}

    \bibliographystyle{apsrev}
    
    \title {Reflection from black holes 
            and space-time topology}
    
    \author{M.Yu.Kuchiev}
    \email[Email:]{kmy@newt.phys.unsw.edu.au}

    \affiliation{ School of Physics, University of New South Wales,
      Sydney 2052, Australia}
    

      \begin{abstract}
        The quantum corrections make the black hole capable of
        reflection: any particle that approaches the event horizon can
        bounce back in the outside world. The albedo of the black hole
        depends on its temperature. The reflection shares physical
        origins with the phenomenon of Hawking radiation; both effects
        are explained as consequences of the singular nature that the
        event horizon exhibits on the quantum level.
       \end{abstract}


    \maketitle

      \label{none}
    
      Intuitively black holes are defined as objects that do not emit
      anything and possess ideal absorption properties. A known
      limitation for this naive picture stems from thermodynamics of
      black holes. The first indication that gravitational fields
      could have entropy came when investigation of Christodoulou
      \cite{christodoulou_1970} of the Penrose process for extracting
      energy from a Kerr black hole showed that there is a quantity
      which could not go down.  Hawking found \cite{hawking_1970} that
      it is proportional to the area of the horizon. Further research
      of Bardeen {\it et al} \cite{bardeen_carter_hawking_1970}
      demonstrated that black holes obey laws similar to the laws of
      thermodynamics.  The next step made by Bekenstein
      \cite{bekenstein_1972} indicated that the area was actually the
      physical entropy. The discovery of the Hawking radiation
      \cite{hawking_1974} and the Unruh process \cite{unruh_1976}
      completed foundation of the thermodynamics of black holes; for a
      recent review see Ref.\cite{Wald_2001}, see also books
      \cite{frolov_novikov_1998,
        chandrasekhar_1993} for comprehensive discussion of black hole
      properties.  The thermodynamics reveals that a black hole has
      the finite temperature and is capable of radiation through the
      Hawking mechanism.
    
      Classically, the metric of a black hole and the equations of
      motion of a probing particle in this metric are described by
      regular functions on the horizon, if considered in appropriate
      coordinates, see e.g.  \cite{misner_thorne_wheeler_1973}. This
      fact allows the incoming particle to cross the horizon smoothly,
      making the black hole an ideal absorber.  However, this work
      shows that black holes are capable to reflect particles from the
      event horizon.  The effect arises due to quantum corrections,
      becoming apparent in the semiclassical approximation. One can
      describe it in terms of the classical action of the probing
      particle that possesses a singularity on the horizon.  The
      singularity, that persists in any coordinate frame, triggers the
      singularity of the wave function that describes the incoming
      particle.  As a result the wave function of the incoming
      particle acquires an unexpected admixture of the outgoing wave.
      It is this admixture that explains the reflective abilities of
      black holes.  It is interesting that the Hawking radiation
      process can also be tracked down to the same origin.  Having
      same basic roots, the two mentioned effects have different
      experimental manifestations because the flux and spectrum of
      reflected particles depend on the flux and spectrum of incoming
      particles, while the radiation depends entirely on properties of
      the black hole.  A consistent quantum treatment of the incoming
      particle agrees with the semiclassical results, revealing also
      an alternative interesting way to account for the quantum
      corrections. It is based on a topological charge that can be
      associated with events that happen with the incoming particle on
      the event horizon.  This charge can be considered as a
      particular topological characteristic of the Schwarzschild
      geometry that becomes important for quantum description of the
      probing particle, being irrelevant at the classical level. The
      topological properties of the space-time of black holes within
      the frames of the classical approximation were discussed by
      Einstein and Rosen \cite{einstein_rosen_1935}, Wheeler
      \cite{wheeler_1955}, and Misner and Wheeler
      \cite{misner_wheeler_1957}.

      
      Consider a black hole with the Schwarzschild geometry
    \begin{equation}
      \label{schw}
      ds^2\!
      = \!-\left(1\!-\! \frac{1}{r}\right)dt^2 + \frac{dr^2}{1-1/r}
    + r^2 (d \theta^2 + \sin^2 \! \theta \,d\varphi^2)\,.
      \end{equation}
      Relativistic units $\hbar=c=1$ supplemented by condition $2Gm =
      1$ imposed on the gravitational constant $G$ and the black hole
      mass $m$ are used. A sphere with the Schwarzschild radius $r=r_g
      \equiv 1$ represents the event horizon.  Let us assume that the
      probing particle is a scalar with the mass $\mu$ and describe
      its motion by the wave function $\Phi$ that satisfies the
      Klein-Gordon equation $g^{\kappa\lambda} \nabla_\kappa \nabla_
      \lambda \Phi=\mu^2 \Phi$, where $\nabla$ is the covariant
      derivative for a given metric. Separating the angular variables
      for the Schwarzschild metric (\ref{schw}) one finds that the
      radial wave function $\phi(r)$ that describes the motion with
      the energy $\varepsilon$ and momentum $L$ satisfies
      \begin{eqnarray}
        \label{phi''}
&&\phi'' + \left(\frac{1}{r}+\frac{1}{r-1} \right) \phi'\\ \nonumber && +
\frac{1}{1-1/r}\left( \frac{\varepsilon ^2}{1-1/r} -\mu^2 
- \frac{L(L+1)} {r^2} \right)  \phi = 0~.
      \end{eqnarray}
      Consider the proper incoming and outgoing waves
      $\phi_\mathrm{in}^{(0)}(r),~\phi_\mathrm{out}^{(0)}(r)$ that
      satisfy
      $\phi_\mathrm{in}^{(0)}(r)=[\phi_\mathrm{out}^{(0)}(r)]^*$ and
      are distinguished by their asymptotic behavior at infinity $
      \phi_\mathrm{in}^{(0)}(r) \rightarrow B \exp [-i(pr
      +\cdots~)\,],~r\rightarrow \infty$, where $p=(\varepsilon^2
      -\mu^2)^{1/2}$ is the momentum, the dots refer to the
      $\mathcal{O}(\ln r/r)$ terms that appear due to long-range
      nature of the gravitational force, and $B$ is a normalization
      factor.  Eq.(\ref{phi''} ) has a regular singularity at the
      horizon, $\phi(r) \rightarrow (r-1)^{\eta},~r\rightarrow 1$.
      Here the index $\eta$ found from Eq.(\ref{phi''}) equals $\eta =
      \mp i\varepsilon$.  For the incoming wave the sign minus here is
      to be chosen, which gives
      \begin{eqnarray}
        \label{horizon}
      \phi_\mathrm{in}^{(0)} (r) = \exp[-i \varepsilon \ln
      (r-1)\,],\quad r\rightarrow1~.
      \end{eqnarray}
      Let us describe the incoming particle in the outside region
      $r>1$ by the wave function
      \begin{eqnarray}
        \label{phi}
\phi_\mathrm{in} (r) = \phi_\mathrm{in}^{(0)}(r) + 
\mathcal{R} \, \phi_\mathrm{out}^{(0)}(r)~.
     \end{eqnarray}
     The first term here, the proper incoming wave, is definitely
     present because we do consider the incoming particle. The
     second term, the proper outgoing wave, is introduced in order to
     allow for the possibility of reflection of the incoming particle
     off the black hole. The magnitude of this, hypothetical at the
     moment, reflection is measured by the coefficient $\mathcal{R}$.
     The unitarity condition requires that $| \mathcal{R} |\le 1$.
     Intuitive arguments based on an experience with classical
     trajectories would indicate that the black hole is incapable of
     reflection, i.e. the reflection coefficient in (\ref{phi}) is
     zero.  However, the quantum corrections do produce the phenomenon
     of reflection, making $|\mathcal{R}|>0$, as verified below.
     
\begin{figure}[b]
\centering
\includegraphics[height=2.5cm,keepaspectratio=true]{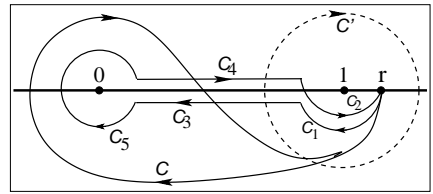}
\caption{\it  
  The complex $r$-plane. To verify Eqs.(\ref{phi}),(\ref{R=}) it is
  sufficient to continue analytically the incoming wave along $C_1$
  and the outgoing wave along $C_2$.  In the semiclassical
  approximation the shift along the contour $C$ transforms the
  incoming wave into the outgoing wave.  The reflection coefficient
  (\ref{R=}) can be reproduced as an eigenvalue of a along shift the
  contour $C'$, which has the topological number $n=1$.}\label{one}
\end{figure}

     Let us consider what happens with the incoming wave in
     Eq.(\ref{phi}) in the inside region $r<1$. The horizon is a
     classically forbidden area  for the outside observer who needs to
     wait an infinite interval of time to see the incoming particle to
     cross the horizon.  However, this fact does not prevent the
     incoming wave from penetrating into the inside region simply
     because this is a {\it stationary} wave that exists during the
     infinite interval of time. This claim is supported by the
     asymptotic behavior of the wave function on the horizon
     (\ref{horizon}). Continuing this wave function analytically from
     the outside region $r>1$ into the inside region $r<1$ over the
     lower semiplane of the complex $r$-plane, see the counter $C_1$
     in Fig.\ref{one}, one finds that this wave exists in the region
     $r<1$, being suppressed compared with the outside region $r>1$ by
     the factor $k = \exp(-\pi\varepsilon)$. We will return to this
     point below.
     \noindent 
     Consider now the singularity of Eq.(\ref{phi''}) at the origin
     $r=0$.  There are two solutions here, the regular $\phi(r)
     \propto 1$, and the singular $ \phi(r) \propto \ln r $.
     Conventional quantum mechanical arguments favor the regular
     boundary condition $\phi(0) = const$, which cannot be satisfied
     by the incoming wave by itself. One needs to introduce its
     complex conjugate, i.e. the proper outgoing wave, and take a
     linear combination of these two waves.  These arguments show that
     the {\it outgoing} wave certainly exists in the inside region
     $r<1$ prompting one to consider what happens with this
     wave on the horizon.  The problem is similar to the one discussed
     above for the incoming wave. The outgoing wave is stationary,
     hence it is able to penetrate into the outside region $r>1$. Its
     analytical continuation along the contour $C_2$ in Fig.\ref{one}
     gives the necessary suppression factor, which again proves be
     equal to $k = \exp(-\pi \varepsilon)$.
     
     We find that the wave function for the incoming particle has the
     form of Eq.(\ref{phi}) that incorporates the {\em expected}
     proper incoming wave, represented by the first term, and the {\em
       surprising} admixture of the proper outgoing wave, the second
     term. The above discussion shows that the reflection arises after
     the horizon is traversed twice, along $C_1$ and $C_2$, each
     crossing introducing a suppression factor $k = \exp(-\pi
     \varepsilon)$.  Hence the reflection coefficient in
     Eq.(\ref{phi}) satisfies
     \begin{eqnarray}
       \label{R=}
       |\mathcal { R }| = k^2 = \exp(-2\pi \varepsilon\,)~.
     \end{eqnarray}
     The admixture of the outgoing wave in Eq.(\ref{phi}) shows that
     black holes are capable of reflection, see Ref.\cite{kuchiev_2003}
     that derives this fact from symmetry conditions. In the presented
     approach the effect can be tracked down to the singularity that
     exists in the wave equation (\ref{phi''}) at the horizon. The
     above arguments can be cast into conventional mathematical terms.
     The singularity $r=1$ is associated with the $2\times 2$
     monodromy matrix $\cal{M}$.  The eigenvalues $\lambda_{1,2}$ of
     this matrix are related to the indexes $\eta = \pm i\varepsilon$,
     $\lambda_{1,2} = \exp(2\pi i \eta) = \exp (\mp 2\pi
     \varepsilon)$.  The absolute value for the reflection coefficient
     in Eq.(\ref{R=}) coincides with the smaller eigenvalue. This
     construction has a topological bearing.  A closed contour on the
     complex $r$-plane around the singularity on the horizon $r=1$ can
     be associated with the topological charge $n$ that counts its
     winding around the point $r=1$ and gives the eigenvalues
     $\exp(\mp 2\pi \varepsilon n)$ for the $n$-th power of the matrix
     $\mathcal{M}$. An interesting part of this construction is that
     the reflective ability of black holes can be associated with
     those events that are distinguished by the topological number
     $n=1$. Continuing the incoming wave along the counter $C'$ with
     $n=1$ in Fig.1 one reproduces the reflection coefficient
     Eq.(\ref{R=}) as an eigenvalue of this transformation.
      
     Alternatively, the effect of reflection can be described in the
     semiclassical approximation, with the help of the classical
     action $S({\bf r},t)$. For the metric (\ref{schw}) the variables
     are separated $S({\bf r},t) = - \varepsilon t+L \varphi +S(r)$,
     where $\varphi$ is the azimuthal angle. The radial part of the
     action $S(r)$ found from the Hamilton-Jacobi classical equations
     of motion $g^{\kappa\lambda}\partial_\kappa S \partial_\lambda S
     = -\mu^2$ reads $S(r) = \mp S_0 (r)$ (see e.g.
     \cite{misner_thorne_wheeler_1973}), where
       \begin{eqnarray}
         \label{S0}
         S_0(r) = \! \int^r \!\!
         \left[ \varepsilon^2 -
        \left(\mu^2+L^2/r^2 \right) 
\left(1-1/r \right) 
         \right]^{1/2} \! \! \frac{dr}{1-1/r}\,.
       \end{eqnarray}
       On the horizon this action possesses the logarithmic singularity
       \begin{equation}
         \label{Ln}
         S_0(r) \simeq \varepsilon \ln(r-1)~, \quad r \rightarrow 1~.
       \end{equation}
       Eqs.(\ref{horizon}),(\ref{Ln}) shows that the wave function on
       the horizon exhibits a semiclassical behavior
       $\phi_\mathrm{in}^{(0)} (r) = \exp[-i S_0(r)\,]$.  Therefore
       the found above suppressing factor $k=\exp(-\pi \varepsilon)$
       that describes the crossing of the horizon can be attributed to
       the semiclassical approximation. Moreover, all arguments
       regarding the behavior of the incoming and outgoing waves in
       Eq.(\ref{phi}) can be reproduced semiclassically. One takes the
       proper incoming wave for $r>1$, then continue it along the
       contour {\it C} in Fig.  \ref{one} (which can be transformed
       into a series of contours {\it C$_1$-C$_5$}) and end up with
       the proper outgoing wave. These purely semiclassical arguments
       supports validity of Eqs.(\ref{phi}),(\ref{R=}). The
       derivation given holds for arbitrary $\mu,\varepsilon,L$.
       However, to allow the particle a classically allowed
       propagation from the horizon towards the infinity the classical
       momentum must be real for $1 \le r < \infty $. To satisfy this
       condition it suffices to have either $L/2 \le \mu$ for
       arbitrary energy $\varepsilon \le \mu$, or $L\le
       (3\sqrt3/2)\,\varepsilon$ for the ultraleativistic case $
       \varepsilon \gg \mu$.

\begin{figure}[htb]
\centering
\includegraphics[height=7cm,keepaspectratio=true]{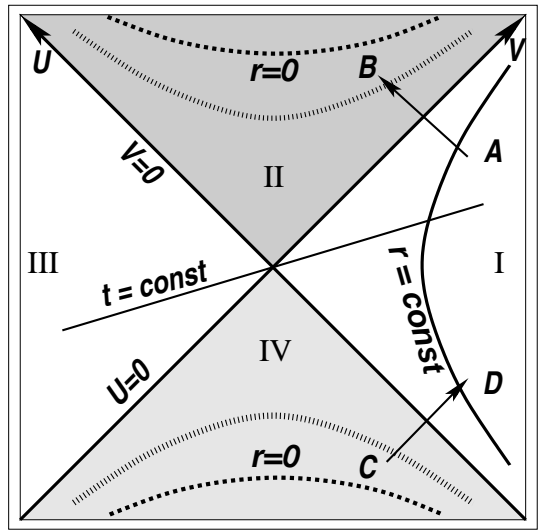}
\caption{\it 
  Kruskal coordinates. Areas {\it I} and {\it III} represent two
  identical copies of the outside region; {\it II, IV} show two inside
  regions. Hyperbolic curves $UV = const$ describe condition
  $r=const$, the dotted curve shows location of $r=0$, the inclined
  straight line presents condition $t=const$. The incoming particle
  follows {\it AB} crossing the horizon $U=0$ and residing in {\it
    II}. The outgoing particle {\it CD} escapes from {\it IV} crossing
  the horizon $V=0$ and coming to {\it I}.  Areas {\it II} and {\it
    IV} are not connected, which ensures classical confinement in {\it
    II}.  The wave finction (\ref{gen}) describes mixing of events
  that correspond to incoming and outgoing classical trajectories
  ({\it AB} and {\it CD}), resulting in phenomena of reflection and
  radiation.  }
\label{two}
\end{figure}

       Thus the reflection ability of black holes arises from the
       logarithmic singularity of the classical action (\ref{Ln}).  It
       is important that this singularity is an invariant property of
       the action that persists even in those coordinates that
       eliminate the singularity of the metric and classical equations
       of motion.  Consider for example convenient Kruskal coordinates
       $U,V$ that are defined by equations
       \begin{eqnarray}
         \label{U}
U/V = -\exp(-t), \quad
U V = (1-r) \exp(r)~.
       \end{eqnarray}
       Fig. \ref{two} shows that there are four regions of the
       space-time, see Ref.\cite{misner_thorne_wheeler_1973} for
       detailed discussion.  Areas {\it I\/} and {\it III\/} represent
       two identical copies of the outside region; {\it II\/}, {\it
         IV\/} show two inside regions. They all are divided by the
       horizon located at $U=0$, or $V=0$.  The incoming particle
       follows the trajectory {\it AB\/} and, crossing the horizon,
       resides in {\it II\/}. There are also the outgoing trajectories;
       following the trajectory {\it CD\/} the particle escapes from
       the inside region {\it IV\/} into the outside world {\it I\/}.
       Areas {\it II\/} and {\it IV\/} are not connected, which
       ensures (classical) confinement of the incoming particle
       trapped in {\it II\/}.  
\noindent 
       In Kruskal coordinates the metric is regular on the horizon,
       also regular are the classical equations of motion that on the
       horizon read simply $V=const$ for $|U|\ll |V| \ll 1$ and
       $U=const$ for $|V|\ll |U|\ll 1$. However, the singularity of
       the action remains intact. It can be conveniently presented in
       terms of the action $\mathcal{S} = -\varepsilon t \mp S_0(r)$,
       where the minus and plus signs are for the incoming and
       outgoing trajectories respectively,
       \begin{eqnarray}
         \label{lnS}
         \mathcal {S} 
          \simeq  
\left\{ 
\begin{array}{r}
-\varepsilon 
\ln (V^2) ~,\quad \quad 
         |U|\ll |V| \ll 1~,
\\         
\varepsilon 
 \ln (U^2)~,\quad \quad
         |V|\ll |U| \ll 1 ~. \end{array}
\right.      
       \end{eqnarray}
       The wave function (\ref{phi}) can be presented in a more
       general form
       \begin{eqnarray}
\label{gen}
|\,\phi_\mathrm{in} \,\rangle = 
|\,\mathrm{in} \,\rangle + \mathcal{R}\,|\,\mathrm{out} \,\rangle~.
       \end{eqnarray}
       Here the variable $t$ is included in the wave function
       $|\,\phi_\mathrm{in} \,\rangle \equiv \exp(-i\varepsilon
       t)\phi_\mathrm{in} (r)$, permitting convenient
       presentation for the proper incoming $|\,\mathrm{in} \,\rangle
       $ and outgoing $ |\,\mathrm{out} \,\rangle $ waves, which in
       the vicinity of the horizon read
       \begin{eqnarray}
         \label{wfK}
\!\!\! |\,\mathrm{in} \,\rangle 
\!=\! \exp[-i \varepsilon \ln(V^2)\,],
~~\,  
|\,\mathrm{out} \, \rangle \!= \!\exp[\,i \varepsilon \ln(U^2)\,].
       \end{eqnarray}
       Since the semiclassical picture is valid on the horizon, one
       can associate the proper incoming wave, i.e. the first term in
       (\ref{gen}), with the incoming trajectories, see {\it AB} in
       Fig. \ref{two}. Correspondingly, the second term, the proper
       outgoing wave, is associated with the outgoing trajectories,
       see {\it CD}. Eq.(\ref{gen}) shows that effects associated
       with these different classical trajectories are mixed in the
       wave function of the incoming particle. 
       
       Notice that the effects that look completely isolated in the
       classical description interfere in the wave function
       (\ref{gen}).  We proved this fact above for the outside region
       {\it I}, where $U<0,~V>0$.  However, it can be argued that this
       form of the wave function remains valid in a more general case
       because the wave functions in (\ref{gen}),(\ref{wfK}) are
     symmetrical under the transformations $U\rightarrow
     -U,~V\rightarrow -V$. These transformations convert an event in
     the outside region {\it I} into an event in one of the inside
     regions {\it II}, or {\it IV}.  Following this line of arguments
     one concludes that quantum description of any event in any
     region, inside or outside, is based on the wave function
     (\ref{gen}) that includes two terms: the proper incoming and
     outgoing waves.

       Consider implications of this fact for a particle that in the
       classical approximation is confined in the inside region.
       Describing its behavior with the help of the wave function
       (\ref{gen}) one concludes that there exists a probability
       $\mathcal{P}_\mathrm{esc} \propto |\mathcal{R}|^2
       =\exp(-\varepsilon/T)$ for this particle to populate the
       outgoing wave and escape into the outside world. We see that on
       the quantum level confinement cannot be absolute, the locked in
       particle can escape into the outside world. Thus the black hole
       creates a flux of outgoing particles escaping from its inside
       region.  It is instructive to consider the black hole behavior
       inside the temperature bath with the temperature equal to the
       Hawking temperature $T=1/(4\pi)$. In this imaginary experiment
       the black hole absorbs the particles produced by the
       temperature bath and emits particles that escape from its
       inside region.  Considering the ratio of the probability to
       escape to the probability to be absorbed
       $\mathcal{P}_\mathrm{abs}$, one finds \footnote{There is a
         finite probability to cross the event horizon
         $\mathcal{P}_\mathrm{cr} = 1-\mathcal{P}$, where
         $\mathcal{P}$ accounts for the reflection (\ref{R=}), but
         since $ \mathcal{P}_\mathrm{esc}$ and $
         \mathcal{P}_\mathrm{pen} $ are both proportional to this
         crossing probability it is canceled out in (\ref{ratio}).}
      \begin{eqnarray}
        \label{ratio}
\mathcal{P}_\mathrm{esc}/ \mathcal{P}_\mathrm{abs} = 
\exp(-\varepsilon/T)~.
      \end{eqnarray}
      This shows that the black hole remains in equilibrium with the
      temperature bath and hence has the temperature $T=1/(4\pi)$, as
      was discovered in Ref.\cite{hawking_1974}, see also discussion
      in Ref.\cite{hartle_hawking_1976}.  The phenomenon of the
      Hawking radiation is usually explained by creation of pairs in
      the gravitational field. As we see, Eq.(\ref{gen}) proposes an
      alternative simple explanation: the radiation happens because
      the wave function of a particle that is confined inside the
      horizon incorporates an admixture of the outgoing wave that
      gives the particle a chance to escape into the outside world.

      Quantum corrections alter the fundamental properties of the
      event horizon.  In the classical approximation this is an area
      that the incoming particle penetrates through smoothly going
      inside, but which becomes impassable in the opposite direction.
      Quantum effects change both these properties. They make possible
      the reflection from the horizon that shows that the horizon
      presents some obstacle for the incoming particle. The reflection
      is the more prominent the lower is the energy, for $\varepsilon
      < T$ the black hole behaves as a mirror, which is surprising.
      Quantum corrections also make possible the Hawking radiation
      that can be described as an escape of the particle from the
      inside region of a black hole, which shows that the horizon is
      transparent for the inside-out transition.
      
      The discussed phenomena follow from the fact that the classical
      action is singular on the horizon \footnote{ Different behavior
      of the action and classical equations of motion bears
      resemblance with the Aharonov-Bohm effect, in which the action
      depends on the magnetic flux, while classical trajectories do
      not.}, making singular the semiclassical wave function and
      ultimately resulting in Eq.(\ref{gen}). There is a topological
      aspect of the quantum problem. It arises because the singularity
      of the wave equation on the horizon can be associated with the
      particular topological charge $n$ that counts winding of the
      trajectory of the probing particle around the horizon on the
      complex $r$-plane. The reflective and radiative abilities of
      black holes are associated with topologically nontrivial
      trajectories with $n=1$. This result makes the topological
      charge an interesting characteristic of the Schwarzschild
      geometry of the space-time.  Generalization of the discussed
      results for the Kerr-Neumann charged rotating black holes is
      given in Ref.\cite{kuchiev3_2003}. It verifies validity of
      Eq.(\ref{gen}) for this case as well, deriving from it the
      phenomena of reflection and Hawking radiation.

Discussions with V.V.Flambaum are appreciated. This work was supported
by the Australian Research Council.


   \end{document}